\documentclass[a4paper,twocolumn,11pt,accepted=2021-11-18]{quantumarticle}
\pdfoutput=1 
\usepackage[utf8]{inputenc}
\usepackage[english]{babel}
\usepackage[T1]{fontenc}
\usepackage{amsmath}
\usepackage{amssymb}
\usepackage{cite}
\usepackage{bbm}
\usepackage{color}
\usepackage{xcolor,hyperref}
\usepackage{graphicx}
\usepackage{tikz}
\usepackage{lipsum}
\newcommand{\naw}[1]{\left(#1\right)}
\newcommand{\ket}[1]{\left|#1\right>}
\newcommand{\bra}[1]{\left<#1\right|}
\newcommand{\av}[1]{\left<#1\right>}

\newcommand{\modu}[1]{\left|#1\right|}
\newcommand{\poisson}[1]{\left\{#1\right\}}
\usepackage{sidecap}
\usepackage{wrapfig}
\usepackage{subfig}

\begin{document}
\title{Groups, Platonic solids and Bell inequalities}
\author{Katarzyna Bolonek-Laso\'n}
\affiliation{Department of Statistical Methods, Faculty of Economics and Sociology University of Lodz, 41/43 Rewolucji 1905 St., 90-214 Lodz,  Poland}
\author{Piotr Kosi\'nski}
\thanks{piotr.kosinski@uni.lodz.pl}
\affiliation{Department of Computer Science, Faculty of Physics and Applied Informatics University of Lodz, 149/153 Pomorska St., 90-236 Lodz, Poland}

\maketitle
\begin{abstract}
The construction of Bell inequalities based on Platonic and Archimedean solids (\emph{Quantum} \textbf{4} (2020), 293) is generalized to the case of orbits generated by the action of some finite groups. A number of examples with considerable violation of Bell inequalities is presented.
\end{abstract}

\section{Introduction}
Since the pioneering Bell paper \cite{Bell} the Bell inequalities became the subject of intensive study (for a review see \cite{Liang},\cite{Brunner}). Their importance stems from the fact that their violation at the quantum level provides the evidence that the quantum theory cannot be viewed as a local realistic theory. Another important notion in physics is that of symmetry. On the formal level various symmetries are described in the framework of group theory. Therefore, it appears natural to study Bell inequalities for the systems described by the sets of states classified by the representations of some groups. This idea has been proposed in the interesting papers by G\"uney and Hillery \cite{Guney},\cite{Guney1} and studied in some detail by the present authors \cite{Bolonek,Bolonek1,Bolonek2,Bolonek3}. Within this approach the states of quantum system entering some Bell inequality form the orbit(s) of a particular unitary representation of some group $G$. The orbit is chosen in such a way as to consist of a disjoint sum of subsets, each forming an orthonormal basis in the space of states. Each such basis provides the spectral decomposition of some observable. Group-theoretical methods allow for simple calculation of quantum bound on the combination of probabilities entering the particular Bell's inequality. Computing classical bound calls for more effort. However, one can use here Fine's theorem \cite{Fine} which considerably reduces the relevant combinatorics.

Recently, Tavakoli and Gisin \cite{Tavakoli} in the very nice paper constructed Bell inequalities  which violations are attained with measurements pointing to the vertices of the Platonic and Archimedean solids. Again we are dealing with the systems exhibiting high degree of symmetry; it is the symmetry which makes the Platonic solids so beautiful. In the present paper we generalize the Tavakoli \& Gisin approach using the tools from elementary group theory. Our starting point is the idea to look at the Platonic and Archimedean solids (as well as some other threedimensional objects) as generic and nongeneric orbits of threedimensional representations of some finite groups (basically, the subgroups of relevant symmetry groups). It appears that this point of view brings some advantages. We obtain an extremally simple formula for quantum value of the combination of correlation functions introduced in \cite{Tavakoli} under the assumption that our quantum system is described by the maximally entangled $SO(3)$ singlet state. We find nice geometric  picture, following from Fine's theorem and group theory, of classical states saturating Bell inequality. 
Moreover, group theory offers also here some simplifications even though they are less spectacular than in the quantum case. A number of examples of considerable violation of Bell inequality is presented.

Our Bell inequalities are for sure not experimentally friendly. As those proposed in Ref.~\cite{Tavakoli} they require many measurement settings, more than necessary for exhibiting the violation of classical bounds. They are also not very effective. Their visibility (i.e.~the resistance to noise), which in our case equals simply the ratio of classical to quantum bounds, is at best the same as for CHSH Bell inequality. In spite of these disadvantages we believe that the study of Bell inequalities for the systems exhibiting symmetry is promising. In opposition to the point of view which has gained recently some popularity \cite{Hossenfelder} we believe that (mathematical) beauty is an important ingredient of any really successful theory or model. This is the case for General Relativity, Weinberg-Salam model or QCD. However, the role of symmetry in the field of Bell inequalities and, more generally, entanglement, is not clear. One can look, for example, for Alice-Bob settings with increasing resistance to noise, improving that of CHSH inequality. This can be done with the help of powerful and efficient algorithms. The progress is, however, moderate. The first improvement of the CHSH value $v=\frac{1}{\sqrt{2}}\approx 0.7071$ appeared in 2008 \cite{Vertesi}, $v\simeq 0.7056$, and demanded 465 settings for both parties. Quite recent improvement \cite{Brierley}, based on 42 settings, yielded $v\simeq 0.7012$. They are not experimentally friendly at all. Moreover, they are obtained with the help of certain algorithms and their meaning is not clear; one would expect the settings optimalizing the visibility to exhibit some kind of symmetry. This question can be put in a wider context. It is well known that there are various manifestations of quantum entanglement which partially overlap. There are some conjectures concerning more detailed description of this overlapping like, for example, the Peres conjecture \cite{Peres}. Vertesi and Brunner disproved it by presenting a counterexample \cite{Vertesi1}. Again, this counterexample has been obtained by applying powerful numerical algorithm and its meaning seems to remain obscure.

Generally speaking, the role of symmetries in entanglement phenomena is the topic worth to be explored. In the subsequent paper we will analyze the Bell inequalities for the Alice and Bob settings exhibiting $"$hidden$"$ symmetry, i.e.~generated by group action but not coinciding with Platonic or Archimedean solids.

\section{The general scheme}

We consider a Bell experiment with Alice and Bob having a number of possible settings characterized by the sets of directions in space, $\poisson{\vec{v}_i}_{i=1}^{N_A}$ and $\poisson{\vec{w}_j}_{j=1}^{N_B}$, respectively. They measure the spin $=\frac{1}{2}$ projections in the $\vec{v}_i$ (Alice) and $\vec{w}_j$ (Bob) directions; the corresponding observables $A_i$ and $B_j$ acquire the values $\pm 1$ depending on whether the projection is positive or negative.

One expects the Bell inequality to be violated the more the larger is the degree of entanglement of the state of the system. Therefore, we choose this state to be the $SO(3)$ singlet
\begin{equation}
\ket{\phi}\equiv\frac{1}{\sqrt{2}}\naw{\ket{\downarrow\uparrow}-\ket{\uparrow\downarrow}}.\label{1}
\end{equation}
It differs from the state $\ket{\phi^+}\equiv\frac{1}{\sqrt{2}}\naw{\ket{\uparrow\uparrow}+\ket{\downarrow\downarrow}}$ used by Tavakoli and Gisin \cite{Tavakoli}; however it is maximally entangled, $\mathrm{Tr}_A\naw{\ket{\phi}\bra{\phi}}=\frac{1}{2}\mathbbm{1}=\mathrm{Tr}_B\naw{\ket{\phi}\bra{\phi}}$ and slightly more convenient from the theoretical point of view since the relevant correlation functions are rotationally invariant. Moreover, most results obtained for $\ket{\phi^+}$ remain valid here (see below).

Simple calculation yields the well-known formula for the correlation function
\begin{equation}\begin{split}
\av{A_iB_j}_\phi\equiv \bra{\phi}\naw{\vec{v}_i\cdot\vec{\sigma}}\otimes\naw{\vec{w}_j\cdot\vec{\sigma}}\ket{\phi}=-\vec{v}_i\cdot\vec{w_j} \label{2}\end{split}
\end{equation}
which is rotationally invariant as expected. 

The Bell inequality we are considering concerns a linear combination of correlation functions
\begin{equation}
\mathcal{B}=\sum_{i=1}^{N_A}\sum_{j=1}^{N_B}c_{ij}\av{A_iB_j}.\label{3}
\end{equation} 
Following \cite{Tavakoli} we adjust the coefficients $c_{ij}$ in such a way as to optimize the quantum bound on $\mathcal{B}$. Therefore, we put 
\begin{equation}
c_{ij}=-\vec{v}_i\cdot\vec{w}_j\label{4}
\end{equation}
and eq.~\eqref{3} takes the form
\begin{equation}
\mathcal{B}=-\sum_{i=1}^{N_A}\sum_{j=1}^{N_B}\naw{\vec{v}_i\cdot\vec{w}_j}\av{A_iB_j}.\label{5}
\end{equation}
In particular, for the singlet state \eqref{1}
\begin{equation}
\mathcal{B}=\sum_{i=1}^{N_A}\sum_{j=1}^{N_B}\naw{\vec{v}_i\cdot\vec{w}_j}^2.\label{6}
\end{equation}
In Ref.~\cite{Tavakoli} the directions $\vec{v}_i$, $\vec{w}_j$ correspond to the vertices of Platonic solids (and Archimedean ones). Therefore, they enjoy a high degree of symmetry. In particular, the most symmetric case corresponds to dual Platonic solids determining Alice and Bob settings. Dual Platonic solids share the same symmetry group. Guided by this example we consider the following general situation. Let $G=\poisson{g_\alpha}_{\alpha=1}^{|G|}$ $(g_1=e)$ be some finite group admitting threedimensional real irreducible representation; some reducible representations can be also considered (see  below). We assume that $\poisson{v_i}_{i=1}^{N_A}$, $\poisson{w_j}_{j=1}^{N_B}$ are two arbitrary orbits of $G$, generated by the representation $D(g)$ under consideration. The orbits of $G$ are obtained by acting with all matrices $D(g)$ on some fixed vectors,
\begin{equation}
\vec{v}_\alpha=D(g_\alpha)\vec{v}\label{7}
\end{equation} 
\begin{equation}
\vec{w}_\beta=D(g_\beta)\vec{w}.\label{8}
\end{equation}
For generic $\vec{v}$ ($\vec{w}$) we get $|G|$ vectors $\vec{v}_\alpha$ ($\vec{w}_\beta$). However, it can happen that the stability subgroup of $\vec{v}$ ($\vec{w}$), $G_v$ ($G_w$) is nontrivial. Then the vectors $\vec{v}_\alpha$ ($\vec{w}_\beta$) are in one-to-one correspondence with the cosets in $^G/_{G_v}$ ($^G/_{G_w}$), i.e. $\vec{v}_\alpha=\vec{v}_{\alpha'}$ ($\vec{w}_\beta=\vec{w}_{\beta'}$) iff $g_\alpha$, $g_{\alpha'}$ ($g_\beta, g_{\beta'}$) belong to the same coset of $^G/_{G_v}$ ($^G/_{G_w}$). Note that $^{|G|}/_{|G_v|}=N_A$, $^{|G|}/_{|G_w|}=N_B$.
 
The orbits consisting of exactly $\vert G\vert$ elements (i.e.~corresponding to the vectors $\vec{v}(\vec{w})$ admitting only trivial stability subgroup) are called generic; the remaining ones are nongeneric.

Consider now the following sum
\begin{align}
\begin{split}
&\sum_{g,g'\in G}\naw{D(g')\vec{v}\cdot D(g)\vec{w}}^2=\\
&=\sum_{g,g'\in G}\naw{\vec{v}\cdot D(g'^{-1})D(g)\vec{w}}^2=\\
&=\sum_{g,g'\in G}\naw{\vec{v}\cdot D((g'^{-1}\cdot g))\vec{w}}^2=\\
&=|G|\sum_{g\in G}\naw{\vec{v}\cdot D(g)\vec{w}}^2=\\
&=|G|\sum_{a,b,c,d=1}^3\sum_{g\in G}v_a w_b v_c w_d D_{ab}(g) D_{cd}(g)
\end{split} \label{9}
\end{align}
where we have changed the summation indices, $g'\rightarrow g'$, $g'^{-1}\cdot g\rightarrow g$.

By applying the orthogonality relations 
\begin{equation}
\sum_{g\in G}\overline{D_{ab}^{(\mu)}(g)}D_{cd}^{(\nu)}(g)=\frac{|G|}{\dim D^{(\mu)}}\delta_{\mu\nu}\delta_{ac}\delta_{bd}\label{10}
\end{equation}
and the normalization conditions $|\vec{v}|=|\vec{w}|=1$ we find
\begin{equation}
\sum_{g,g'\in G}\naw{D(g')\vec{v}\cdot D(g)\vec{w}}^2=\frac{1}{3}|G|^2.\label{11}
\end{equation}
Now, any term of the sum defining $\mathcal{B}$, eq.~\eqref{6}, enters the left-hand side of \eqref{11} $|G_{v}|\cdot|G_w|$ times. Therefore, one finds finally the following expression for $\mathcal{B}$ for the maximally entangled state \eqref{1}
\begin{equation}
\mathcal{B}=\frac{1}{3}N_A\cdot N_B. \label{12}
\end{equation}
This is an extremally simple formula in terms of the Alice and Bob numbers of settings. It is important to point out that the only assumption we made is that $G$ generates the orbits of Alice and Bob; $G$ does not have to be the full symetry group of either orbit; the latter is much stronger assumption. For example, the symmetric group $S_4$ is the symmetry group of the regular tetrahedron which is some nongeneric orbit of threedimensional representation of $S_4$; however, for judicious choice of initial vector one can also obtain octahedron which has larger symmetry group $S_4\times S_2$.

We shall now show that the bound \eqref{12} is optimal. To this end let us note that, according to the eqs.~\eqref{2}$\div$\eqref{5}
\begin{equation}
\mathcal{B} = \langle\hat{\mathcal{B}}\rangle
\end{equation}
 with the Bell operator defined by
 \begin{widetext}
 \begin{align}
 \hat{\mathcal{B}}\equiv -\sum_{i=1}^{N_A}\sum_{j=1}^{N_B}\naw{\vec{v}_i\cdot\vec{w}_j}\naw{\vec{v}_i\cdot\vec{\sigma}\otimes\vec{w}_j\cdot\vec{\sigma}}=
-\frac{1}{\vert G_v\vert}\cdot\frac{1}{\vert G_w\vert}\sum_{i=1}^{\vert G\vert}\sum_{j=1}^{\vert G\vert}\sum_{k, l, m=1}^3 (\vec{v}_i)_k(\vec{v}_i)_l(\vec{w}_j)_k(\vec{w}_j)_m\sigma_l\otimes\sigma_m.\label{12b}
\end{align}

Consider the expression
\begin{align}
\begin{split}
v_{kl}\equiv\sum_{i=1}^{\vert G\vert}(\vec{v}_i)_k(\vec{v}_i)_l=\sum_{i=1}^{\vert G\vert}\sum_{m,n=1}^3D_{kn}(g_i)(\vec{v})_n D_{ml}(g_i^{-1})(\vec{v})_m.\label{12c}\end{split}
\end{align}\end{widetext}

It is easy to check that, for any $g\in G$,
\begin{equation}
\sum_{m,n=1}^3 D_{km}(g)v_{mn}D_{nl}(g^{-1})=v_{kl}.\label{12d}
\end{equation}
Now, $D$ is irreducible and Schur's lemma implies
\begin{equation}
v_{kl}=V\delta_{kl}.\label{12e}
\end{equation}
Analogously,
\begin{equation}
w_{kl}\equiv\sum_{j=1}^{\vert G\vert}(\vec{w}_j)_k(\vec{w}_j)_l=W\delta_{kl}.\label{12f}
\end{equation}
Eqs.~\eqref{12b}, \eqref{12e} and \eqref{12f} imply
\begin{equation}
\hat{\mathcal{B}}=-\frac{1}{\vert G_v\vert\cdot\vert G_w\vert}VW\cdot\sum_{l=1}^3(\sigma_l\otimes\sigma_l).\label{12g}
\end{equation}
On the other hand, by virtue of eqs.~\eqref{12e}, \eqref{12f} we find
\begin{equation}
\sum_{i=1}^{\vert G\vert}\sum_{j=1}^{\vert G\vert}(\vec{v}_i\cdot\vec{w}_j)^2=3VW.\label{12h}
\end{equation}
Taking into account that $\sum\limits_{l=1}^3(\sigma_l\otimes\sigma_l)$ has the eigenvalues 1 and -3 we conclude that the maximal expectation value of $\hat{\mathcal{B}}$ is given by eq.~\eqref{6}.

Note that the above reasoning remains valid even if the orbits $\{v_i\}$, $\{w_j\}$ correspond to different groups and/or different threedimensional irreducible representations. However, eq.~\eqref{12} is obtained from eq.~\eqref{6} provided we are working with the same group and representation.

Consider now the classical bound $C$ on $\mathcal{B}$. Its determination is slightly more involved. According to Fine's theorem \cite{Fine} it is sufficient to compute $\mathcal{B}$, eq.~(\eqref{5}), for all deterministic responses of Alice and Bob and pick the largest one \cite{Tavakoli}, \cite{Bolonek}.
Therefore, we obtain \cite{Tavakoli}
\begin{small}\begin{align}
\begin{split}
\mathcal{B}\leq C&\equiv \max\limits_{\substack{A_1,\ldots,A_{N_A}\in\{\pm 1\}^{N_A}\\ B_1,\ldots,B_{N_B}\in\{\pm 1\}^{N_B}}}\naw{-\sum_{j=1}^{N_B}B_j \sum_{i=1}^{N_A}A_i\naw{\vec{v}_i\cdot\vec{w}_j}}\\
&=\max\limits_{A_1,\ldots,A_{N_A}\in\{\pm 1\}^{N_A}}\naw{\sum_{j=1}^{N_B}\modu{\sum_{i=1}^{N_A}A_i\naw{\vec{v}_i\cdot\vec{w}_j}}}.\end{split}\label{13}
\end{align}\end{small}
 The bound \eqref{13} can be effectively computed provided $N_A$ and $N_B$ are not too large.
 
 In order to get more insight into the structure of classical bound let us rewrite $C$ in the form
 \begin{equation}
 C=\max\limits_{\substack{A_1,\ldots,A_{N_A}\in\{\pm 1\}^{N_A}\\ B_1,\ldots,B_{N_B}\in\{\pm 1\}^{N_B}}}\naw{\sum_{i=1}^{N_A}A_i\vec{v}_i}\cdot\naw{\sum_{j=1}^{N_B}B_j\vec{w}_j}\label{14}
 \end{equation}
 where we have redefined $A_i\rightarrow-A_i$. Therefore, $C$ is expressed in terms of scalar products of vectors which are linear combinations, with the coefficients $\pm 1$, of the elements of Alice (Bob) orbit. The set of such vectors may be classified in group-theoretical terms. Let us consider a particular combination (for definiteness, we consider Alice orbit)
 \begin{equation}
 \sum_{i=1}^{N_A}A_i\vec{v}_i,\quad A_i=\pm 1.\label{15}
 \end{equation}
 Let $N_A^+$ ($N_A^-$) be the number of $+1$ ($-1$) coefficients entering \eqref{15}, $N_A^++N_A^-=N_A$. The group $G$ acts as some subgroup of permutations of the vectors $\vec{v}_i$. Therefore, by acting with $G$ on the combination \eqref{15} one obtains the set of vectors of the same form with fixed $N_A^{\pm}$. One concludes that the set of vectors \eqref{15} with $N_A^{\pm}$ fixed decomposes into disjoint sum of $G$ orbits. Accordingly, the set of all vectors of the above form is also a disjoint sum of such orbits. In order to determine their form it is sufficient to consider the configurations with $N_A^+ \leq N_A^-$; the remaining ones are obtained by applying space inversion. Once the numbers $N_A^+\leq N_A^-$ are chosen one picks a particular combination \eqref{15}. In order to find the orbit obtained by acting with the elements of $G$ on this combination one has to determine the stability subgroup. The latter contains those elements of $G$ which permute $\vec{v}_i$'s with the same coefficients. However, in general it is larger; this is because the initial orbit $\poisson{\vec{v}_i}_{i=1}^{N_A}$ may contain the subset of vectors which are linearly dependent with the coefficients $\pm 1$ (this is, for example, the case for the cube which contains four pairs of opposite vectors). The same procedure may be applied to Bob's orbit. Once this is completed one can compute $C$ from eq.~\eqref{14}.  

 Contrary to the quantum bound the above described procedure does not seem to be more effective than direct evaluation of the bound $C$ from eq. \eqref{13} by brute force. However, it provides a nice geometrical picture of classical configurations which still allows for some simplifications.
 
First, note the following. The sums in eqs.~\eqref{13} and \eqref{14} run over all different vectors; in other words the indices $i$ and $j$ refer to the elements of cosets $G/G_v$ and $G/G_w$, respectively. Let us call $C'$ the expression which is the counterpart of the right hand sides of eqs.~\eqref{13} and \eqref{14} obtained under the assumption that the sums run over all elements of $G$ even if the initial vectors correspond to nongeneric Alice's and/or Bob's orbits (cf.~eqs.~\eqref{7} and \eqref{8}),
\begin{small}\begin{equation}
C'=\max\limits_{\substack{A_1,\ldots,A_{\vert G\vert}\in\{\pm 1\}^{\vert G\vert}\\ B_1,\ldots,B_{\vert G\vert}\in\{\pm 1\}^{\vert G\vert}}}\naw{-\sum_{\beta=1}^{\vert G\vert}B_\beta\sum_{\alpha=1}^{\vert G\vert}A_\alpha\naw{\vec{v}_\alpha\cdot\vec{w}_\beta}}.
\end{equation}\end{small}
Any vector $\vec{v}_i$ ($\vec{w}_j$) enters $C'$ $\vert G_v\vert$ ($\vert G_w\vert$) times. Assume that the maximum on the right hand side is attained for some choice of $A_\alpha$'s and $B_\beta$'s. Note that 
\begin{align}\begin{split}
\naw{\sum_{\alpha=1}^{\vert G \vert} A_\alpha\vec{v}_\alpha}&\cdot\naw{\sum_{\beta=1}^{\vert G \vert}B_\beta\vec{w}_\beta}=\\
&=\naw{\sum_{\alpha=1}^{\vert G_v\vert}A_\alpha}\vec{v}_1\cdot\naw{\sum_{\beta=1}^{\vert G\vert}B_\beta\vec{w}_\beta}+\\
&+\naw{\sum_{\alpha=\vert G_v\vert+1}^{\vert G\vert}A_\alpha\vec{v}_\alpha}\cdot\naw{\sum_{\beta=1}^{\vert G\vert}B_\beta\vec{w}_\beta}.\label{15a}\end{split}
\end{align}
We easily conclude that maximum is attained provided $A_1=A_2=\ldots=A_{\vert G_v\vert}=\text{sgn}(\vec{v}_1\cdot\sum_{j=1}^{N_B}B_j\vec{w}_j)$. This reasoning applies as well to other vectors $\vec{v}_i$ and the elements of the second orbit. Therefore, one finds immediately 
\begin{equation}
C=\frac{C'}{\vert G_v\vert\cdot\vert G_w\vert}\label{15b}
\end{equation}
and it remains to compute $C'$. Inspecting eqs.~\eqref{13} and \eqref{14} we easily conclude that not all combinations of $A$'s and $B$'s need to be taken into account. For example, if all $A$'s are equal, $\sum\limits_{i=1}^{\vert G\vert}A_i\vec{v}_i$ is invariant under the action of $G$ since the latter only permutes $v_i$'s and, due to the irreducibility of the representation under consideration, $\sum\limits_{i=1}^{\vert G\vert}A_i\vec{v}_i=0$. The same conclusion concerns $B$'s. In general, only those combinations of $A$'s (and $B$'s) should be considered for which $\sum\limits_{i=1}^{\vert G\vert}A_i\vec{v}_i$ ($\sum\limits_{j=1}^{\vert G\vert}B_j\vec{w}_j$, respectively) has nonvanishing projection on the relevant representation. This observation may be formalized as follows. Let $D^{(\mu)}$ be some irreducible representation of finite group $G$; we assume that the matrices $D^{(\mu)}$ are explicitly unitary. We are interested in the expressions of the form $\sum\limits_{i=1}^{\vert G\vert}A_i(\vec{v}_i\cdot\vec{w}_j)$ entering eq.~\eqref{13}. Let $\tilde{D}^{(\mu)}$ be the representations equivalent to $D^{(\mu)}$. Since $\tilde{D}^{(\mu)}$ is, in general, not explicitly unitary, the orthogonality and completeness relations read \cite{Hamermesh}
\begin{equation}
\sum_{j=1}^{\vert G\vert}\tilde{D}^{(\mu)}_{\alpha\beta}(g_j)\tilde{D}_{\gamma\delta}^{(\nu)}(g_j^{-1})=\frac{\vert G\vert}{\dim\tilde{D}^{(\mu)}}\delta_{\mu\nu}\delta_{\alpha\delta}\delta_{\beta\gamma}\label{15c}
\end{equation}
\begin{equation}
\sum_\mu\frac{\dim D^{(\mu)}}{\vert G\vert}\text{Tr}(\tilde{D}^{(\mu)}(g)\tilde{D}^{(\mu)}(g'^{-1}))=\delta_{gg'}.\label{15d}
\end{equation}
For any irreducible representation $\tilde{D}^{(\nu)}$ (chosen arbitrarily except that it coincides with $\tilde{D}^{(\mu)}$ for $\mu=\nu$) we define
\begin{align}
\begin{split}
v_{\alpha\beta,m}^{(\nu)}&=\sum_{i=1}^{\vert G\vert}\tilde{D}_{\alpha\beta}^{(\nu)}(g_i^{-1})(\vec{v}_i)_m=\\
&=\sum_{i=1}^{\vert G\vert}\sum_{n=1}^3\tilde{D}_{\alpha\beta}^{(\nu)}(g_i^{-1})D_{mn}^{(\mu)}(g_i)(\vec{v})_n.\label{15e}\end{split}
\end{align}
Due to the orthogonality condition $v_{\alpha\beta,m}^{(\nu)}=0$ for $\nu\neq\mu$. Eq.~\eqref{15e} implies the inverse relation
\begin{align}
\begin{split}
(\vec{v}_i)_m&=\sum_\nu\sum_{\alpha,\beta}\frac{\dim D^{(\nu)}}{\vert G\vert}\tilde{D}_{\beta\alpha}^{(\nu)}(g_i)v_{\alpha\beta,m}^{(\nu)}=\\
&=\frac{\dim D^{(\mu)}}{\vert G\vert}\sum_{\alpha,\beta}\tilde{D}_{\beta\alpha}^{(\mu)}(g_i)v_{\alpha\beta,m}^{(\mu)}.\label{15f}\end{split}
\end{align}
Therefore,
\begin{align}\begin{split}
&\sum_{i=1}^{\vert G\vert}A_i(\vec{v}_i\cdot\vec{w}_j)=\\
&=\frac{\dim D^{(\mu)}}{\vert G\vert}\sum_{m,n=1}^3\sum_{\alpha,\beta}\overline{A}_{\beta\alpha}D_{mn}^{(\mu)}(g_j)\overline{v_{\alpha\beta,m}^{(\mu)}}(\vec{w})_n \label{15g}\end{split}
\end{align}
where
\begin{equation}
A_{\beta\alpha}\equiv\sum_{i=1}^{\vert G\vert}A_i\tilde{D}_{\beta\alpha}^{(\mu)}(g_i).\label{15h}
\end{equation}
Eq.~\eqref{15g} simplifies combinatorics considerably. In the present context we are dealing with real threedimensional representations. Therefore, eqs.~\eqref{15g} and \eqref{15h} tell us that there are only nine combinations of $A_i$'s which enter the Bell inequality \eqref{13}. In the examples described below we consider mainly the symmetric group $S_4$. Then: (\textit{i}) $\vert G\vert=24$ so the number of possibilities reduces considerably and (\textit{ii}) $\tilde{D}^{(\mu)}$ can be chosen to consist of matrices having 0, $\pm$1 as the matrix elements which makes $A_{\beta\alpha}$ particularly simple.
 
 \section{Some examples}
Let us consider some examples. In all cases considered below the Alice and Bob orbits are represented either by Platonic (tetrahedron, octahedron, cube) or Archimedean (cuboctahedron, truncated octahedron) solids. Their symmetry groups are $S_4$ for tetrahedron and $O_h$ for the remaining solids. The relevant threedimensional irreducible representations of both groups are explicitly described in Appendix B. The sets $\{\vec{v}_i\}_{i=1}^{N_A}$, $\{\vec{w}_j\}_{j=1}^{N_B}$ of Alice and Bob settings are the orbits of the symmetry group generated by the action of the representation matrices on some carefully selected initial vectors. As it has been already noticed above it can happen that the relevant orbit may be obtained by acting on the initial vector $\vec{v}$ with the elements of smaller group. This is, for example, the case if the stability group $G_v$ of the vector $\vec{v}$ is a normal subgroup of the symmetry group $G$; then the orbit is generated by the action of the quotient group $G/G_v$. In fact, all solids we consider, except the cube, are obtained by the action of $S_4$. Summarizing, the orbits we are considering, read:
\begin{itemize}
\item tetrahedron - generated by $S_4$
\item octahedron - generated by $S_4$ or $O_h$
\item cube - generated by $O_h$
\item cuboctahedron - generated by $S_4$ or $O_h$
\item truncated octahedron - generated by $S_4$ or $O_h$.
\end{itemize}

The algorithm described in the previous section can be directly applied to all pairs of the above solids by setting $G=S_4$ or $G=O_h$; the only exception is the pair tetrahedron-cube because the corresponding orbits are generated by different groups, $S_4$ and $O_h$, respectively ($O_h$ is not the symmetry of tetrahedron while $S_4$ does not generate cube). Still this case can be quite easily dealt with (see below).

Let us note that our orbits are, in general, nongeneric. Therefore, they correspond to nontrivial stability subgroups of $S_4$. As a result, the initial vectors must be carefully selected. To this end we have found all eigenvectors corresponding to the eigenvalue 1, of the matrices belonging to the threedimensional representation of $S_4$, described in Appendix B. Each eigenvector is shared by all elements belonging to the corresponding stability subgroup. There are eight (including phase $\pm$1 arbitrariness) normalized eigenvectors shared by six group elements. They form two regular tetrahedrons in dual position; each tetrahedron can be obtained from any of its vertices by acting with $S_4$ group elements.

There are six (again - including the phase $\pm 1$ freedom) eigenvectors shared by three group elements. They form the regular octahedron. There are one-parameter families of normalized eigenvectors shared by two group elements (they obviously include 14 eigenvectors described above). We have selected one particular eigenvector which generates, via $S_4$ group action, the cuboctahedron.
All other threedimensional vectors give rise to generic orbits consisting of 24 elements. Again we selected one which yields, under the action of $S_4$, the truncated octahedron.

\begin{figure}[!htp]
\centering\includegraphics[scale=0.45]{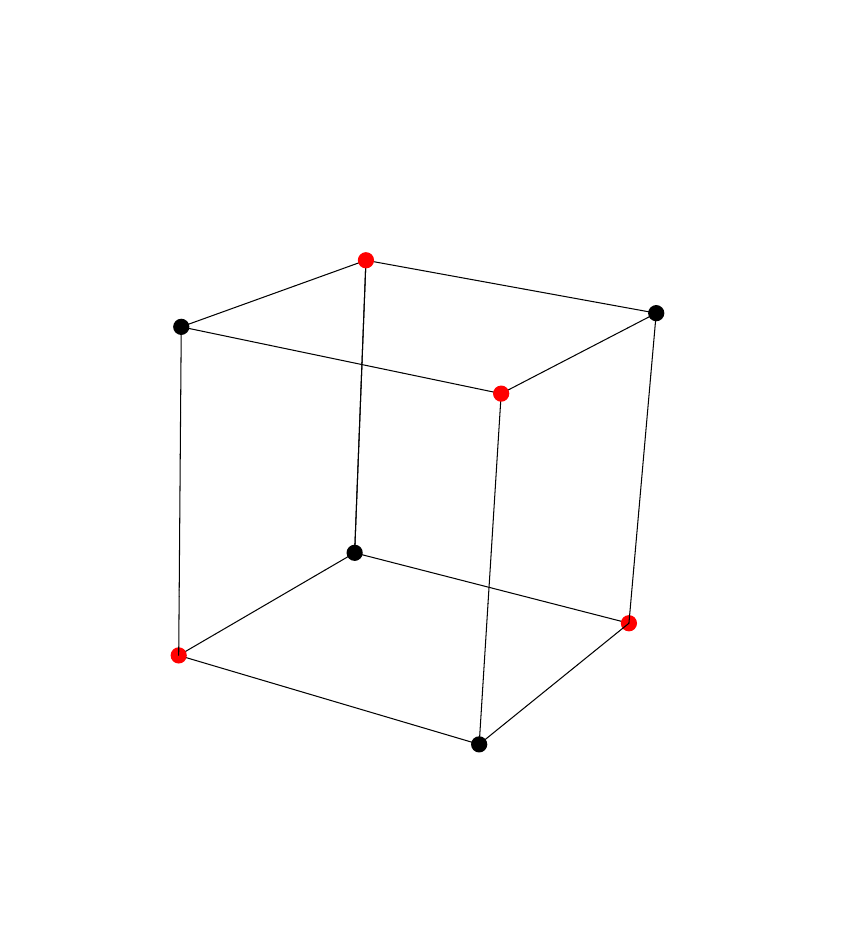}
\caption{Cube as two (red and black) tetrahedrons in dual position.}\label{Figure1}
\end{figure}

Finally, cube cannot be obtained by the action of $S_4$ alone; one needs the whole $O_h$ group. It is easy to see that one can start with the vector generating, via $S_4$, the regular tetrahedron and then, by acting with $-I$ (cf.~Appendix B) on the latter, add the dual tetrahedron. This construction is depicted on Figure \ref{Figure1}.
The details concerning all solids discussed here are collected in Appendix A. The solids described in Appendix A were used to from 11 pairs of orbits determining the sets of Alice and Bob settings. They are written out in Table \ref{t1} where the corresponding classical and quantum bounds are also presented.
\begin{widetext}
\begin{table}
\begin{tabular}{|l|c|c|}
\hline
Alice - Bob & Classical bound & Quantum bound\\
\hline
cuboctahedron - tetrahedron & 13.0639 & 16\\
cuboctahedron - octahedron & 16.9706 & 24\\
cuboctahedron - cube & 26.1279 & 32\\
cuboctahedron - cuboctahedron & 40 & 48\\
truncated octahedron - tetrahedron & 24.7871 & 32\\
truncated octahedron - octahedron & 42.9325 & 48\\
truncated octahedron - cube & 49.5742 & 64\\
truncated octahedron - cubocthedron & 75.8947 & 96\\
truncated octahedron - truncated octahedron & 160 & 192\\
tetrahedron - octahedron & 6.9282 & 8\\
cube - octahedron & 13.8564 & 16\\
\hline
\end{tabular}
\caption{Classical and quantum bounds for various combinations of Alice vs. Bob orbits.}\label{t1}
\end{table}\end{widetext}

\begin{figure}[!h]
\subfloat[cuboctahedron - \\tetrahedron]{\includegraphics[width=0.23\textwidth]{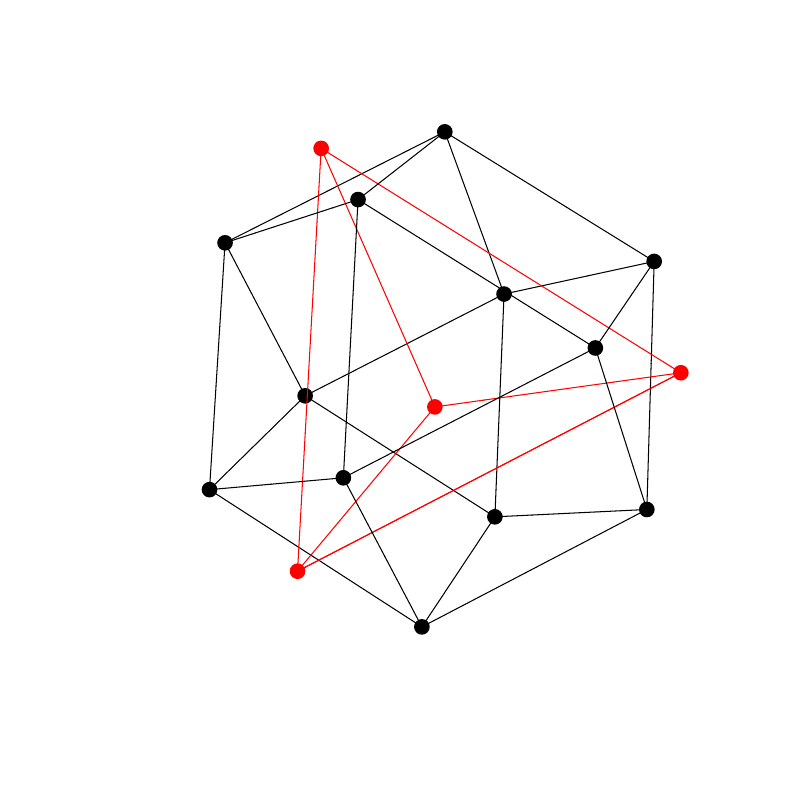}}
\subfloat[cuboctahedron -\\ octahedron]{\includegraphics[width=0.23\textwidth]{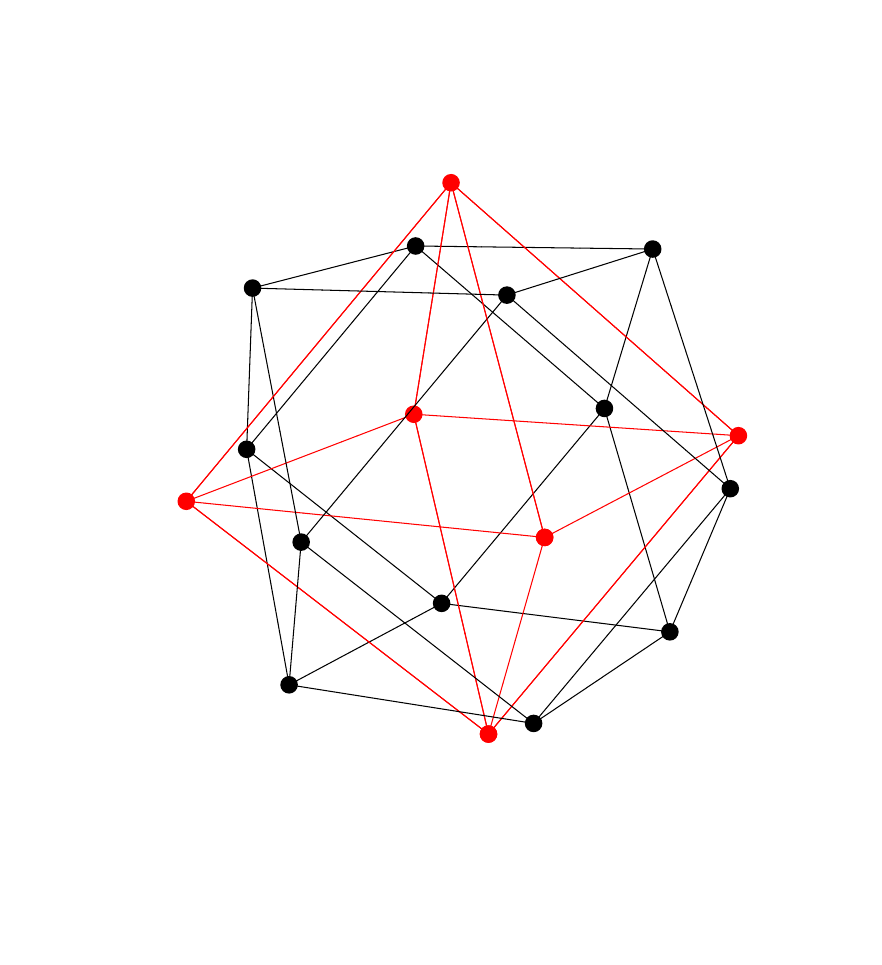}}\\
\subfloat[truncated octahedron -\\ tetrahedron]{\includegraphics[width=0.23\textwidth]{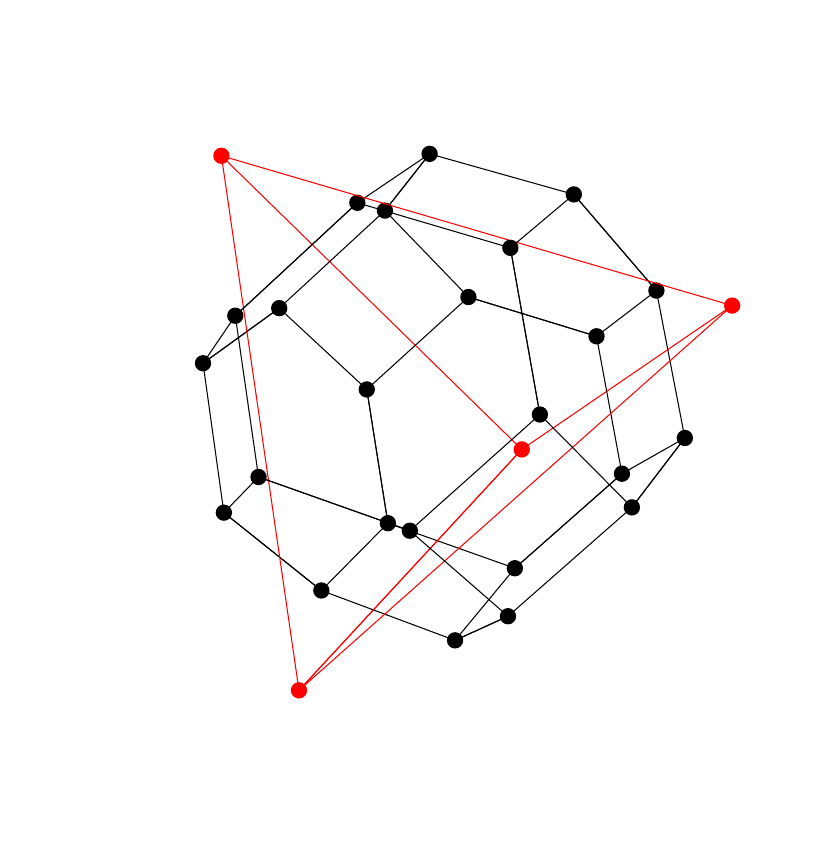}}
\subfloat[truncated octahedron - octahedron]{\includegraphics[width=0.23\textwidth]{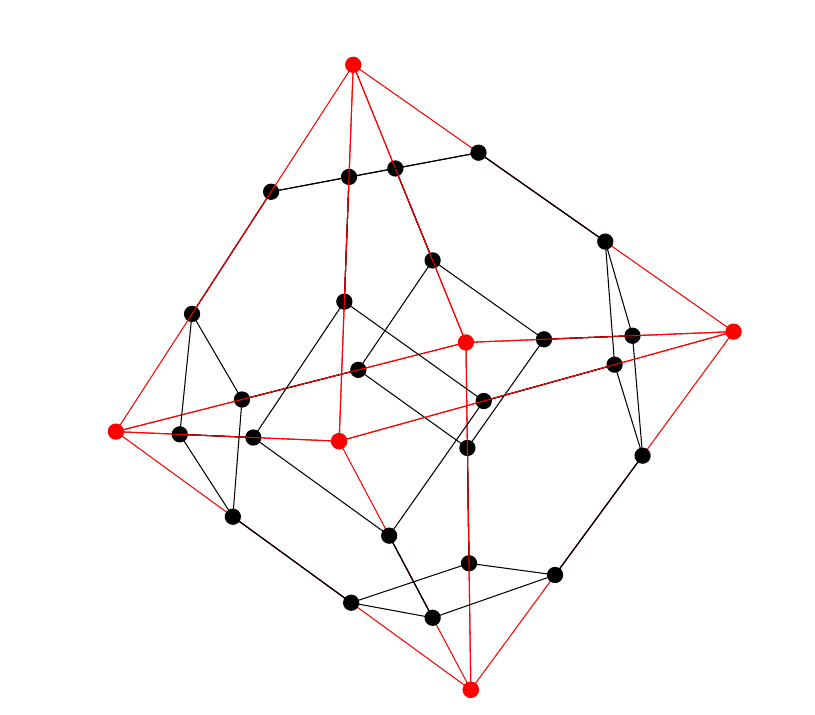}}
\caption{Examples of Alice (black) and Bob (red) settings.}
\label{Fig2}
\end{figure}

Additionally, some examples of Alice and Bob settings are depicted on Figure \ref{Fig2}.

%\begin{widetext}
%\end{widetext}

The content of Table \ref{t1} calls for a number of comments.
\begin{itemize}
\item[1)] Quantum bounds are obtained with the help of eq.~\eqref{12}. In 8 cases the common group generating both orbits can be either $S_4$ or $O_h$; in the remaining 3 cases, which involve cube as one of the orbits, the common group must be $O_h$. However, eq.~\eqref{12} is valid for any choice of the generating group provided it is common for both orbits. Let us note in passing that even the case tetrahedron-cube can be dealt with group-theoretical method. The cube obtained by acting with elements of $O_h$ on some initial vector may be viewed as two tetrahedrons in dual position (cf.~Fig.~\ref{Figure1}). Now cube-cube configuration leads, according to eq.~\eqref{12}, to the quantum bound $\frac{64}{3}$; the classsical bound appears to be the same. Obviously, for tetrahedron-cube settings one finds one half of this result, i.e.~$\frac{32}{3}$.
\item [2)] The classical bounds, presented in Table \ref{t1}, are approximate values. In fact, it follows from eqs.~\eqref{13}, \eqref{14} and the forms of $\vec{v}_i$'s and $\vec{w}_j$'s, presented in Appendix A, that they are expressible as rational functions of a number of square roots of some integers. However, the expressions for exact values are not very transparent, in particular as far as their magnitude is concerned. Therefore, we decided to present approximate values computed with the help of Mathematica.
\item[3)] Computing the classical bound involves maximization over all choices of $A_i$'s and $B_j$'s, $A_i=\pm 1$ $B_j=\pm 1$ (cf.~eqs.~\eqref{13} and \eqref{14}). In principle, this is done by brute force calculations. However, as explained in the previous section, the latter can be considerably simplified. To this end one rewrites the linear combinations of the relevant scalar product, entering second eq.~\eqref{13}, in the form \eqref{15g}. It appearss that only nine combinations of $A_i$'s, given by eq.~\eqref{15h}, enter. By choosing $\tilde{D}(g)$ (representation of $S_4$ equivalent to the initial threedimensional irreducible one) in the form involving only the matrices with matrix elements 0, $\pm 1$ we find the particularly simple form of these nine combinations of $A_i$'s; it is presented in Appendix C. Note that we can work with $S_4$ representation only; for the configurations involving cube we apply the trick of decomposing the cube into two tetrahedrons.
\item[4)] As we have explained in Section II, the sets of vectors $\sum\limits_{i=1}^{N_A}A_i\vec{v}_i$, entering the classical bound \eqref{13}, may be classified in group-theoretical terms (this is, actually, the basis for the simplification described in p.~3). In fact, the sums with the fixed numbers $N_A^+$, $N_A^-$  of positive and negative coefficients, respectively, decompose into disjoint sums of group orbits. Few examples are presented on Figures \ref{F3} and \ref{F4}. Figure \ref{F3} refers to all orbits generated by tetrahedron settings.

The second example concerns the "classical" vectors $\sum\limits_{i=1}^{N_A}A_i\vec{v}_i$, $\sum\limits_{j=1}^{N_B}B_j\vec{w}_j$ saturating the classical bound \eqref{14} for the setting: tetrahedron (Alice)-octahedron (Bob). The orbits are generated by the vectors:

Alice: $\vec{V}\equiv\sum\limits_{i=1}^4A_i\vec{v}_i=\vec{v}_1-\vec{v}_2+\vec{v}_3+\vec{v}_4=(\frac{2}{3},-\frac{4\sqrt{2}}{3},0)$

Bob: $\vec{W}\equiv\sum\limits_{j=1}^{6}B_j\vec{w}_j=-\vec{w}_1+\vec{w}_2-\vec{w}_3+\vec{w}_4-\vec{w}_5+\vec{w}_6=(\frac{2}{\sqrt{3}},-4\sqrt{\frac{2}{3}},0)$
and are presented on Figure \ref{F4}.

\begin{figure}[!h]
\centering
\subfloat[two trivial orbits (red point) corresponding to $N_A^+=0,4$; the vertices of black tetrahedron represent Alice's settings]
{\includegraphics[width=0.3\textwidth]{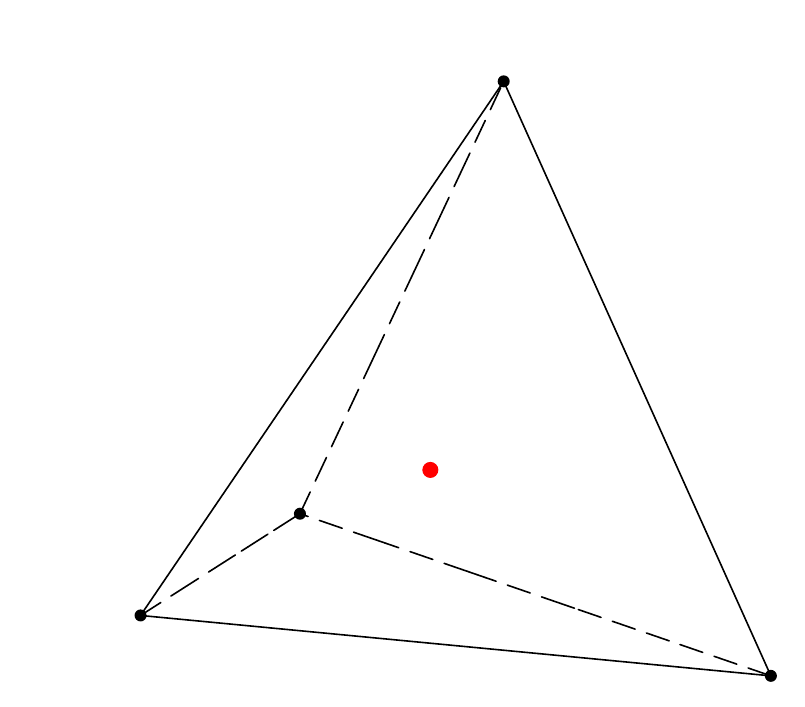}}\quad
\subfloat[the orbit (red octahedron) corresponding to $N_A^+=2$]{
\includegraphics[width=0.3\textwidth]{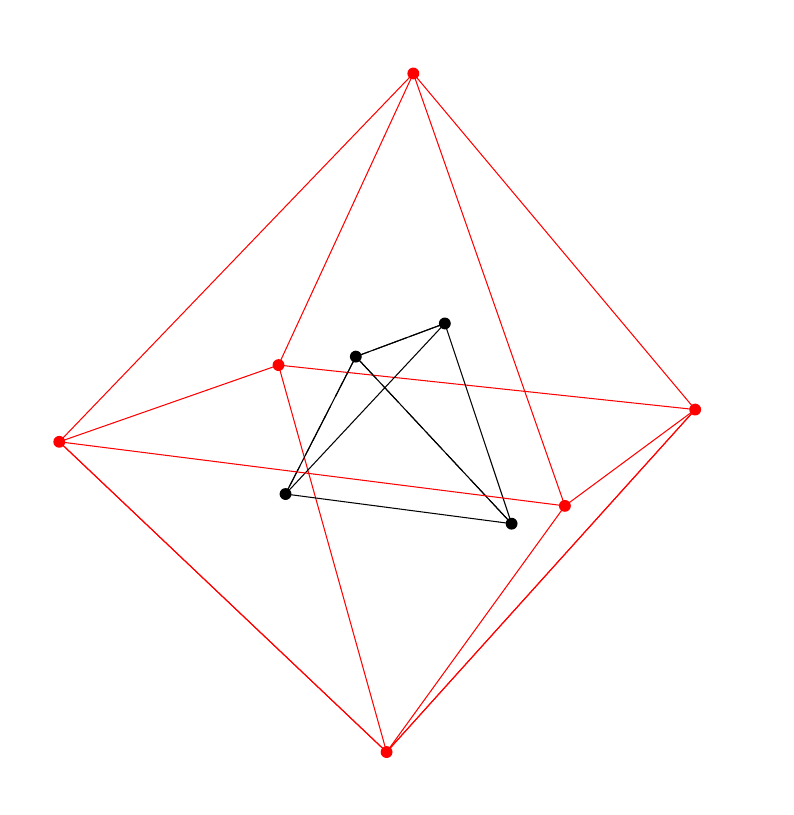}}\quad
\subfloat[two orbits (red and blue tetrahedrons) corresponding to $N_A^+=1,3$]{
\includegraphics[width=0.3\textwidth]{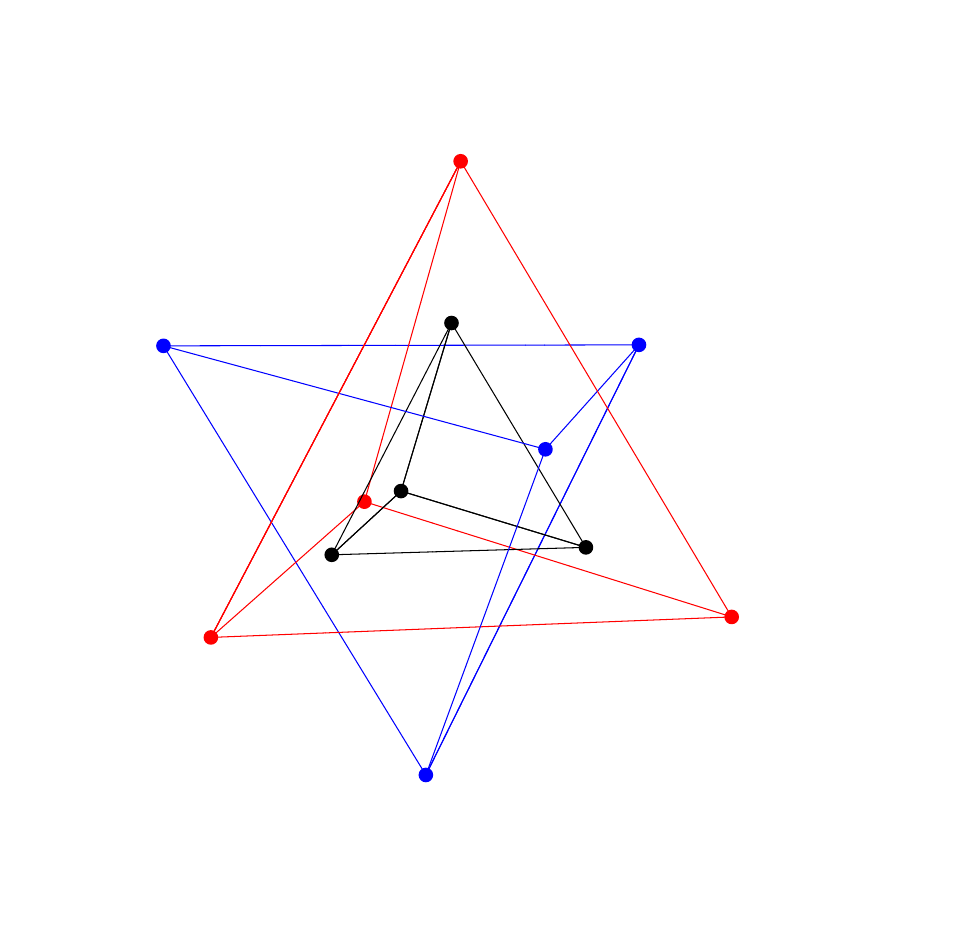}}
\caption{The "classical" orbits for tetrahedron.}\label{F3}
\end{figure}

\begin{figure}[t]
\centering
\subfloat[the initial Alice's settings (magenta) and the $S_4$ orbit generated by the vector $\vec{V}$]{\includegraphics[width=0.18\textwidth]{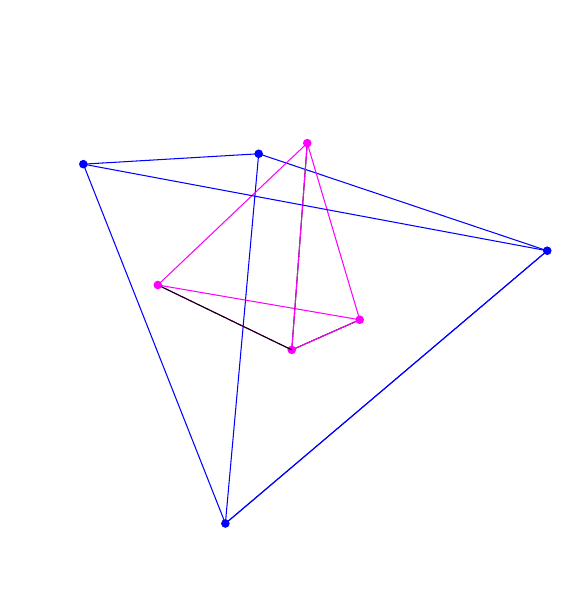}}\quad
\subfloat[the initial Bob's settings (black) and two $S_4$ orbits generated by the vectors $\vec{W}$ and $-\vec{W}$]{\includegraphics[scale=0.25]{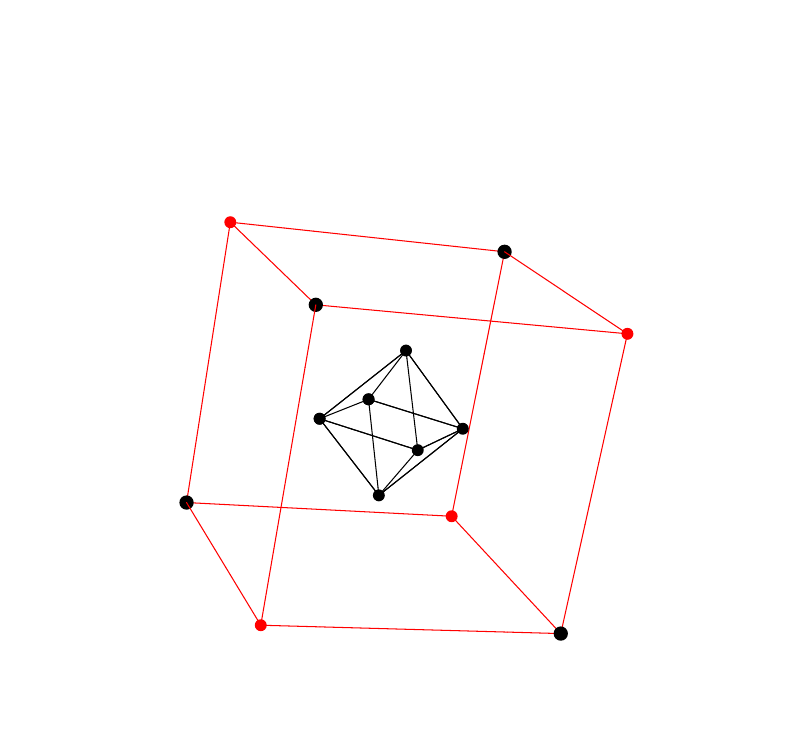}}\\
\subfloat[the $S_4$ orbits generated by $\vec{V}$, $\vec{W}$ and $-\vec{W}$]{\includegraphics[width=0.22\textwidth]{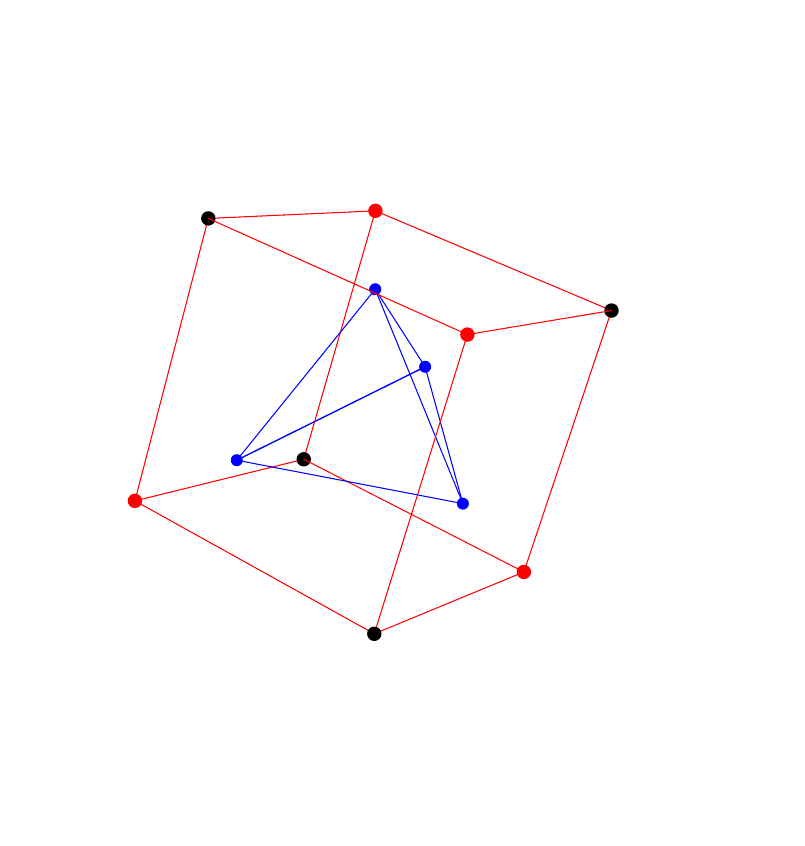}}\quad
\subfloat[the initial Alice's and Bob's settings and the $S_4$ orbits generated by $\vec{V}$, $\vec{W}$ and $-\vec{W}$]{\includegraphics[width=0.23\textwidth]{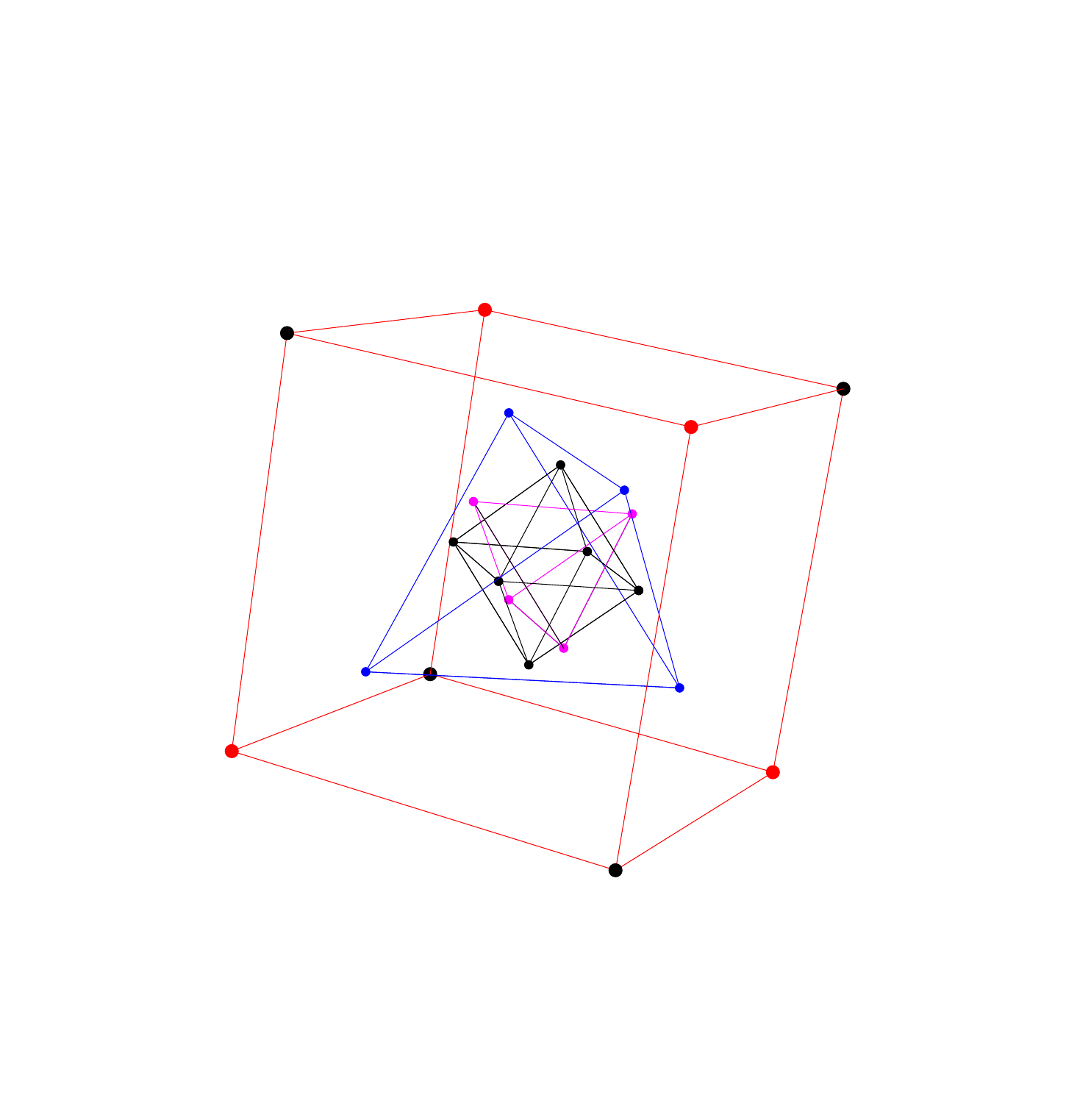}}
\caption{The $S_4$ orbits generated by Alice's (magenta and blue) and Bob's (black and red) vectors. Figs.~4c and 4d present the relative position of the Alice and Bob orbits generated by "classical" vectors.}\label{F4}\end{figure}

\item[5)] One can compare some of the above results with those obtained by Tavakoli and Gisin \cite{Tavakoli}. For example, for cube-octahedron case we find the same values. On the other hand in the tetrahedron-octahedron case the classical bound obtained here is better than the one quoted in \cite{Tavakoli} - 7.82. This is because the relative positions of solids differ: in \cite{Tavakoli} they are chosen to be given by "PolyhedronData" of Mathematica's software while here they are in position resulting from group-theoretical consideration. 

The best quantum-to-classical bounds ratio is obtained for the octahedron-cuboctahedron configuration; it equals 1.4142$\simeq\sqrt{2}$, the same as for CHSH inequality. It is likely that in this case the set of all Alice and Bob settings can be somehow decomposed into the subsets leading to the original CHSH inequality (cf.~Sec.~8 of Ref.~\cite{Tavakoli}).
\end{itemize}
In all cases considered above the shape and relative position of Alice's and Bob's orbits is constrained by group theory. For example, if both orbits are tetrahedrons they must be regular and either coincide or be dual to each other. The reason for that is that they are nongeneric orbits with stability group of order 6. This fact strongly reduces the number of possibilities. Computing the classical and quantum bounds we find no violation of Bell inequality. To become more flexible we should look for a group for which the generic orbits have four vertices, i.e.~a group of order 4. Now, the regular tetrahedron is generated by cyclic subgroup of $S_4$ of order four, generated by cyclic permutation (1234).

\begin{figure}
\begin{center}\includegraphics[width=0.4\textwidth]{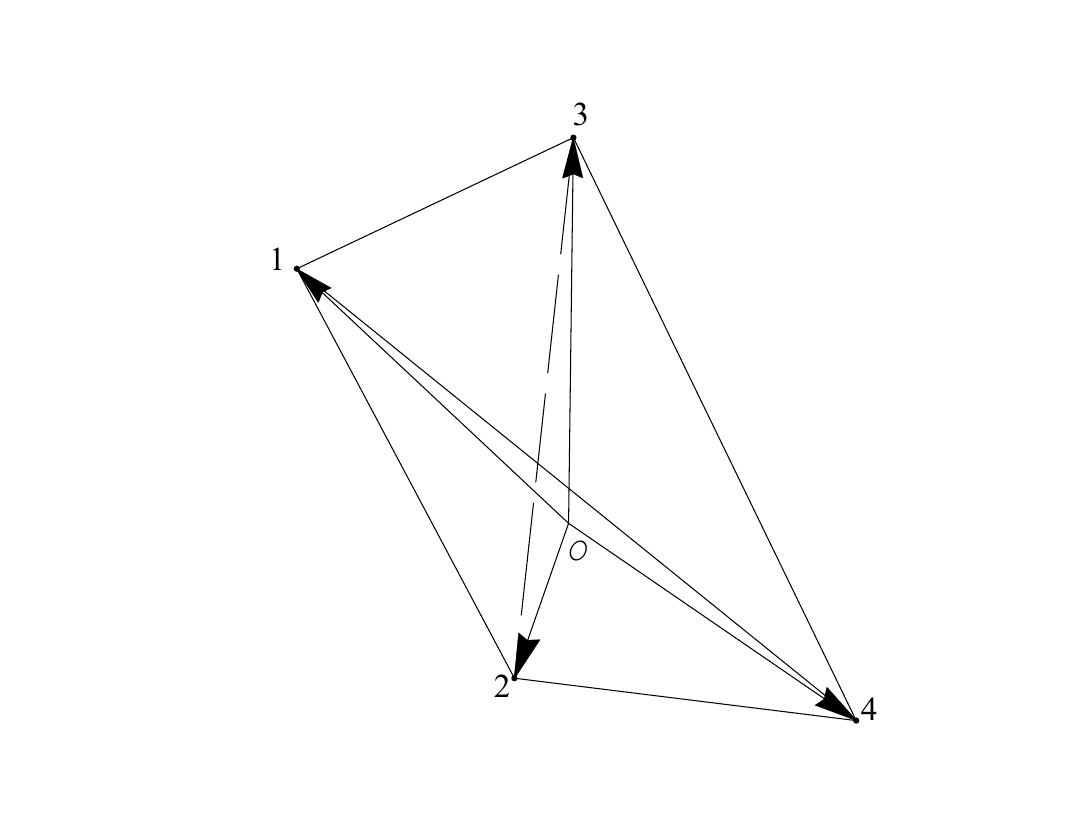}\end{center}
\caption{The generic orbit of $Z_4$; $\measuredangle 1O3=\measuredangle 2O4=\varphi$, $\measuredangle 1O2=\measuredangle 1O4=\measuredangle 2O3=\measuredangle 3O4=\psi$.}
\label{Fig4}
\end{figure}

 So let us take $G=Z_4$ to be cyclic group of order four. Its threedimensional representation can be obtained from that of $S_4$ by subducing it to $Z_4$. In the suitably chosen basis it takes the form (for simplicity we write simply $g$ instead of $D(g)$):
\begin{align}
\begin{split}
& e=\mathbbm{I},\quad g_1\equiv g=\left(\begin{array}{ccc}
-1 & 0 & 0\\
0 & 0 & 1\\
0 & -1 & 0
\end{array}\right),\\
& g_2\equiv g^2=\left(\begin{array}{ccc}
1 & 0 & 0\\
0 & -1 & 0\\
0 & 0 & -1
\end{array}\right), \\
& g_3\equiv g^3=\left(\begin{array}{ccc}
-1 & 0 & 0\\
0 & 0 & -1\\
0 & 1 & 0
\end{array}\right).
\end{split}\label{16}
\end{align}
This representation is reducible as the sum of one- and twodimensional ones. The generic orbits consist of four vertices. Taking
\begin{equation}
\vec{v}_1=\left(\begin{array}{c}
a\\
b\\
c
\end{array}\right ),\quad a^2+b^2+c^2=1\label{17}
\end{equation} 
and denoting $\vec{v}_{k+1}=g_k\vec{v}_1$, $k=1,2,3$, one finds 
\begin{equation}
\vec{v}_2=\left(\begin{array}{c}
-a\\
c\\
-b
\end{array}\right ),\quad \vec{v}_3=\left(\begin{array}{c}
a\\
-b\\
-c
\end{array}\right ), \quad \vec{v}_4=\left(\begin{array}{c}
-a\\
-c\\
b
\end{array}\right ).\label{18}
\end{equation}
Note that $\vec{v}_1\cdot\vec{v}_2=\vec{v}_1\cdot\vec{v}_4=\vec{v}_2\cdot\vec{v}_3=\vec{v}_3\cdot\vec{v}_4=-a^2\equiv\cos\psi$, $\vec{v}_1\cdot\vec{v}_3=\vec{v}_2\cdot\vec{v}_4=a^2-b^2-c^2=\cos\varphi$, $(\vec{v}_2-\vec{v}_4)\cdot(\vec{v}_1-\vec{v}_3)=0$. The resulting orbit is depicted on Fig.~\ref{Fig4}. 
In particular, taking $a^2=b^2=c^2=\frac{1}{3}$ one obtains the regular tetrahedron. 

Let us assume that the Alice orbit is the regular tetrahedron (say $a=b=c=\frac{1}{\sqrt{3}}$) while Bob's orbit corresponds to the vector $\vec{w}$ with arbitrary $a$, $b$, $c$ obeying $a^2+b^2+c^2=1$. First we determine the quantum bound. Eq.~\eqref{12} cannot be applied directly since the representation is reducible. Actually, the orthogonality relations are valid for the representations irreducible in the complex domain. $Z_4$ is abelian so each representation irreducible in the complex domain is onedimensional. In fact $g_1=g$ (and, consequently, all $g_{{\alpha}}$, ${\alpha}=1,2,3$) can be diagonalized using the unitary matrix
\begin{equation}
U=\left(\begin{array}{ccc}
1 & 0 & 0\\
0 & \frac{1}{\sqrt{2}} & \frac{1}{\sqrt{2}}\\
0 & \frac{i}{\sqrt{2}} & -\frac{i}{\sqrt{2}}
\end{array}\right )\label{19}
\end{equation}
so that ($\tilde{g}_\alpha\equiv U^+g_\alpha U$, $\alpha=1,2,3$)

\begin{equation}
\tilde{g}\equiv U^+gU=\left(\begin{array}{ccc}
-1 & 0 & 0\\
0 &  i & 0\\
0 & 0 & -i
\end{array}\right ).\label{20}
\end{equation}
Let $e_\mu(\tilde{g}_\alpha)$, $\mu=1,2,3$,  be the character corresponding to the first, second and third row of $\tilde{g}_\alpha$, respectively. Then any matrix $\tilde{g}_\alpha$ can be written as
\begin{equation}
\naw{\tilde{g}_\alpha}_{ij}=\sum_{\mu=1}^3e_\mu(\tilde{g}_\alpha)\delta_{i\mu}\delta_{j\mu}. \label{21}
\end{equation}
Using eq.~\eqref{21} and the orthogonality relations
\begin{equation}
\sum_{g_\alpha\in Z_4}\overline{e_\mu(\tilde{g}_\alpha)}e_\nu(\tilde{g}_\alpha)=4\delta_{\mu\nu}\label{22}
\end{equation}
one finds
\begin{align}\begin{split}
&\sum_{g_\alpha,g_\beta\in Z_4}\naw{g_\alpha\vec{v},g_\beta\vec{w}}^2=4\sum_{g_\alpha\in G}\naw{\vec{v},g_\alpha\vec{w}}^2=\\
&=4\sum_{g_\alpha\in Z_4}\modu{\naw{\tilde{\vec{v}},\tilde{g}_\alpha\tilde{\vec{w}}}}^2=16\sum_{i=1}^3\modu{\tilde{v}_i}^2\modu{\tilde{w}_i}^2 \label{23}\end{split}\end{align}

where $\tilde{\vec{v}}\equiv U^+\vec{v}$, $\tilde{\vec{w}}\equiv U^+\vec{w}$. Now, $v_i=\frac{1}{\sqrt{3}}$, $i=1,2,3$ and by virtue of eq.~\eqref{19}, $\modu{\tilde{v}_i}=\frac{1}{\sqrt{3}}$. Therefore, eq.~\eqref{23} yields $\frac{16}{3}$ as the quantum bound. As far as the classical bound is concerned one may either classify all orbits consisting of the vectors of the form \eqref{15} and use eq.~\eqref{14} (again quite nice geometrical picture emerges) or use directly eq.~\eqref{13}.

The final result reads 
\begin{equation}
C=\frac{1}{\sqrt{3}}\max\naw{16|a|,8(\modu{b}+\modu{c})}.\label{24}
\end{equation} 
Minimal value of $C$ is acquainted, for example, for $b=\pm\frac{2}{\sqrt{5}}$, $c=0$; then 
\begin{equation}
C=\frac{16}{\sqrt{15}}<\frac{16}{3}\label{25}
\end{equation}
and the Bell inequality is violated.

\section{Summary}
Tavakoli and Gisin presented a very nice picture relating various settings for Alice and Bob, leading to the violation of Bell inequalities, to the geometry of Platonic solids. We have shown here that one can take as a starting point the symmetry groups of Platonic solids or, more precisely, the groups generating these solids as orbits of their threedimensional real representations. This point of view has some advantages. First, it allows the generalization to various groups and generic and nongeneric orbits (including, for example, in unified way the Archimedean solids).

Second, it leads to extremally simple expression for the quantum value of $\mathcal{B}$, given by eq.~\eqref{5}, in terms of the number of Alice's and Bob's settings only (eq.~\eqref{12}). Third, it provides a nice picture of the configurations (eq.~\eqref{15}) entering the formula \eqref{14} for the classical bound which simplifies considerably the brute force calculations of the latter. It should be also noted that the group-theoretical approach determines some natural relative positions of the solids determining Alice's and Bob's settings.
 
An important point is that the group-theoretical point of view offers the possibility of concerning more general configurations of Alice's and Bob's settings. It should be stressed that while the nongeneric orbits with large stability subgroups represent regular (Platonic or Archimedean) solids, this is, generically, no longer the case for larger orbits. There, in order to obtain the regular solid, one has to select carefully the initial vector. Therefore, we may consider configurations which are characterized by a kind of "hidden" symmetry: they are generated by group action but do not look "symmetric" according to the standards of Euclidean geometry. Moreover, we have more freedom in fixing their relative orientation even though it is correlated with their shape. We plan the systematic study of Bell inequalities for such more general settings in subsequent publications.

Finally, let us show that the choice of $\ket{\phi}$ instead of $\ket{\phi^+}$, used in Ref.~\cite{Tavakoli}, does not influence the results. $\ket{\phi^+}$ belongs to the triplet representation of the rotation group; consequently, the relevant correlation function is no longer rotationally invariant
\begin{equation}
\bra{\phi^+}(\vec{v}_i\cdot\vec{\sigma})\otimes(\vec{w}_j\cdot\vec{\sigma})\ket{\phi^+}=\vec{v}_i\cdot\mathcal{I}_y\vec{w}_j\label{26}
\end{equation}  
where $\mathcal{I}_y$ is the reflection in $x-z$ plane. Instead of $\mathcal{B}$ defined by eq.~\eqref{5} we consider the following one
\begin{equation}
\mathcal{B}=\sum_{i=1}^{N_A}\sum_{j=1}^{N_B}\naw{\vec{v}_i\cdot\mathcal{I}_y\vec{w}_j}\av{A_iB_j}.\label{27}
\end{equation} 
In particular, for the $\ket{\phi^+}$ state we get 
\begin{equation}
\mathcal{B}=\sum_{i=1}^{N_A}\sum_{j=1}^{N_B}\naw{\vec{v}_i\cdot\mathcal{I}_y\vec{w}_j}^2.\label{28}
\end{equation}
Now, repeating the previous reasoning we find 
\begin{widetext}\begin{align}
\begin{split}
&\sum_{g,g'\in G}\naw{D(g)\vec{v}\cdot\mathcal{I}_y D(g')\vec{w}}^2=\sum_{a,b,c,d,e,f=1}^3 v_b v_e w_c'w_f'\sum_{g\in G}D(g)_{ab}D(g)_{de}\sum_{g'\in G}\naw{\mathcal{I}_yD(g')\mathcal{I}_y}_{ac}\naw{\mathcal{I}_yD(g')\mathcal{I}_y}_{df}=\\
& =\sum_{a,b,c,d,e,f=1}^3 v_bv_ew_c'w_f'\frac{|G|^2}{9}\delta_{ad}\delta_{be}\delta_{ad}\delta_{cf}=\frac{|G|^2}{3}
\end{split}\label{29}
\end{align}\end{widetext}
because $D(g')$ and $\mathcal{I}_yD(g')\mathcal{I}_y$ are equivalent representations and $\vec{w}'=\mathcal{I}_y\vec{w}$.

As far as classical bound is concerned both formulae, eq.~\eqref{5} and \eqref{27}, give the same results if at least one of the orbits (i.e.~defining Alice's and/or Bob's settings) is invariant under reflection in $x-z$ plane. This is, for example, the case for all pairs of Platonic solids considered in \cite{Tavakoli} which lead to Bell inequality violation.

\bibliographystyle{plain}

\onecolumn\newpage
\appendix
\section{Appendix}
Below we present all solids which form the $G$ - orbits appearing in the text. The description applies both to Alice's and Bob's orbits; in the latter case we should merely make the replacement $\vec{v}\rightarrow\vec{w}$.
\begin{figure}[!h]
\subfloat[the tetrahedron]{\centering\includegraphics[width=0.3\textwidth]{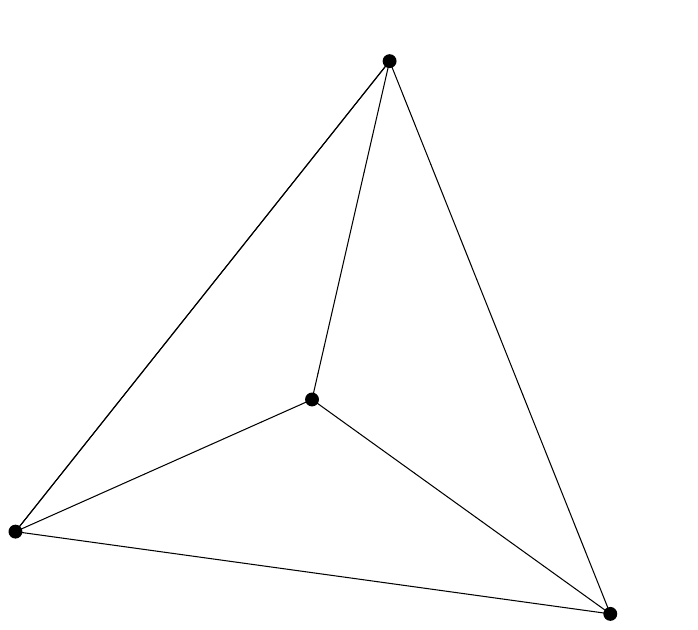}}
\subfloat[the octahedron]{\includegraphics[width=0.3\textwidth]{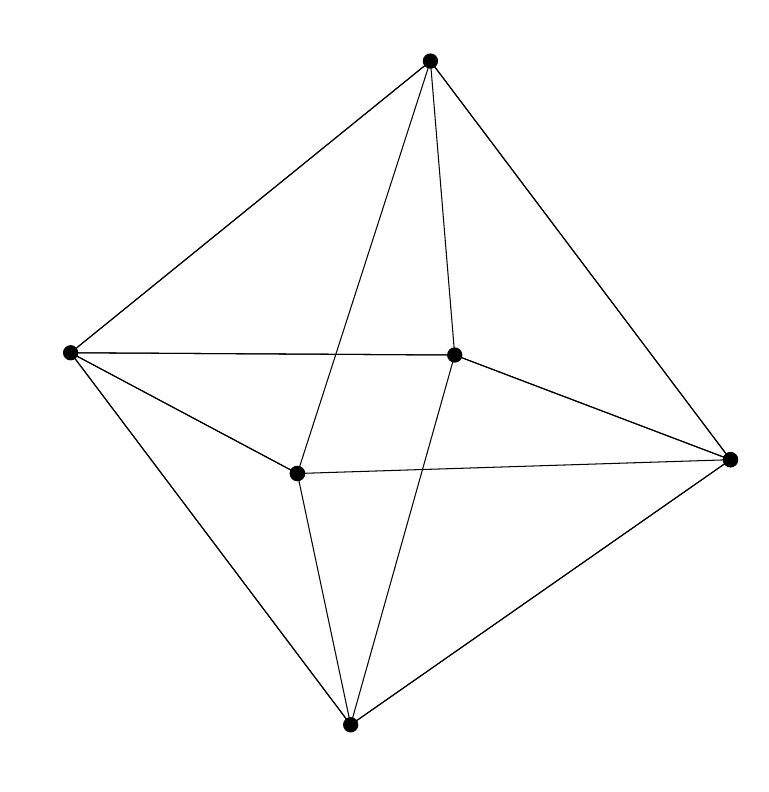}}
\subfloat[the cube]{\includegraphics[width=0.33\textwidth]{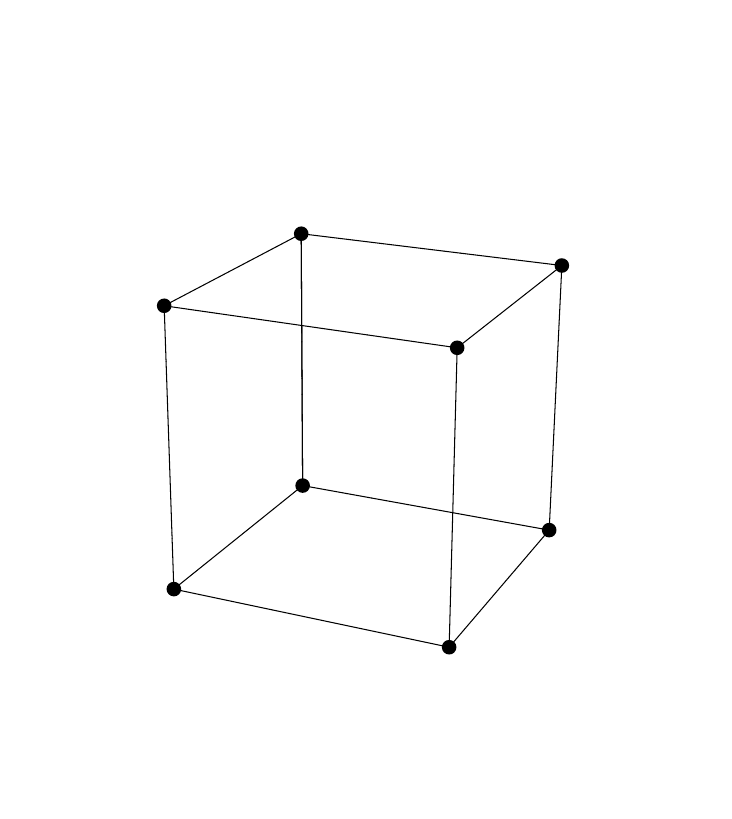}}\\
\centering
\subfloat[the cuboctahedron]{\includegraphics[width=0.3\textwidth]{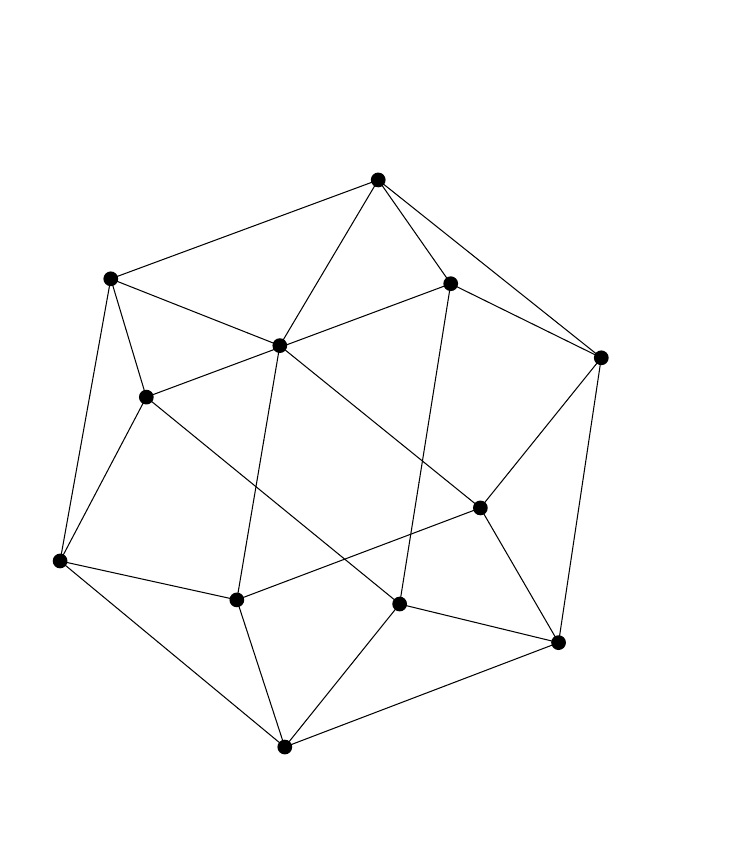}}
\subfloat[the truncated octahedron]{\includegraphics[width=0.3\textwidth]{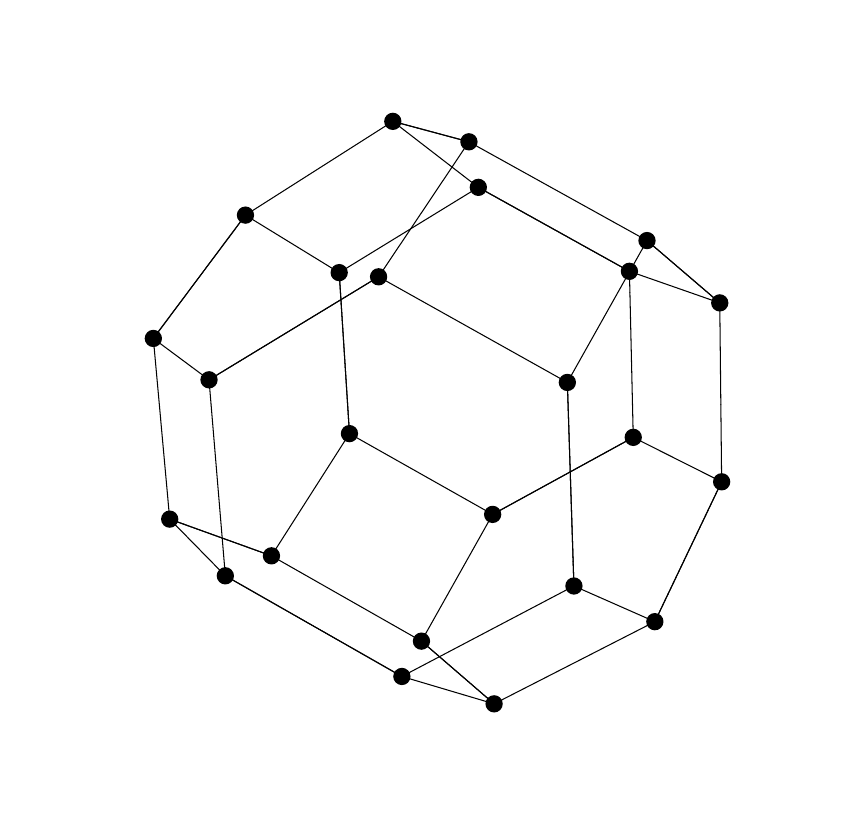}}
\caption{Solids which form the $S_4$-orbits or $O_h$-orbits.}
\end{figure}
\begin{enumerate}
\item The tetrahedron:
\begin{itemize}
\item[] Symmetry group: $S_4$
\item[] Orbit generating group: $S_4$ (or $Z_4$)
\item[] The initial vector: $\vec{v}_1=(1,0,0)$
\item[] The vertices: 
\begin{displaymath}\begin{array}{ll}
v_1=(1,0,0), & v_2=\naw{-\frac{1}{3},-\frac{\sqrt{2}}{3},\sqrt{\frac{2}{3}}}\\
v_3\naw{-\frac{1}{3},\frac{2\sqrt{2}}{3},0}, & v_4\naw{-\frac{1}{3},-\frac{\sqrt{2}}{3},-\sqrt{\frac{2}{3}}}. \end{array}\end{displaymath}
\end{itemize}
\item The octahedron:
\begin{itemize}
\item[] Symmetry group: $O_h$
\item[] Orbit generating group: $S_4$ or $O_h$
\item[] The initial vector: $v_3=\naw{\frac{1}{\sqrt{3}},\sqrt{\frac{2}{3}},0}$
\item[] The vertices:
\begin{displaymath}\begin{array}{ll}
v_1=\naw{-\frac{1}{\sqrt{3}},\frac{1}{\sqrt{6}},\frac{1}{\sqrt{2}}}, &
v_2=\naw{\frac{1}{\sqrt{3}},-\frac{1}{\sqrt{6}},-\frac{1}{\sqrt{2}}},\\ 
v_3=\naw{\frac{1}{\sqrt{3}},\sqrt{\frac{2}{3}},0}, & v_4=\naw{-\frac{1}{\sqrt{3}},-\sqrt{\frac{2}{3}},0},\\
v_5=\naw{-\frac{1}{\sqrt{3}},\frac{1}{\sqrt{6}},-\frac{1}{\sqrt{2}}}, & v_6=\naw{\frac{1}{\sqrt{3}},-\frac{1}{\sqrt{6}},\frac{1}{\sqrt{2}}}. \end{array}\end{displaymath} 
\end{itemize}

\item The cube:
\begin{itemize}
\item[] Symmetry group: $O_h$
\item[] Orbit generating group: $O_h$
\item[] The initial vector: $v_1=\naw{1,0,0}$
\item[] The vertices:
\begin{displaymath}\begin{array}{ll}
v_1=(1,0,0), & v_2=\naw{-\frac{1}{3},-\frac{\sqrt{2}}{3},\sqrt{\frac{2}{3}}},\\
v_3=\naw{-\frac{1}{3},\frac{2\sqrt{2}}{3},0}, & v_4=\naw{-\frac{1}{3},-\frac{\sqrt{2}}{3},-\sqrt{\frac{2}{3}}},\\
v_5=(-1,0,0), & v_6=\naw{\frac{1}{3},\frac{\sqrt{2}}{3},-\sqrt{\frac{2}{3}}},\\
v_7=\naw{\frac{1}{3},-\frac{2\sqrt{2}}{3},0}, & v_8=\naw{\frac{1}{3},\frac{\sqrt{2}}{3},\sqrt{\frac{2}{3}}}.\end{array}\end{displaymath}
\end{itemize}
\item The cuboctahedron:
\begin{itemize}
\item[] Symmetry group: $O_h$
\item[] Orbit generating group: $S_4$ or $O_h$
\item[] The initial vector: $v_1=\naw{-\sqrt{\frac{2}{3}},\frac{1}{\sqrt{3}},0}$
\item[] The vertices:
\begin{displaymath}\begin{array}{ll}
v_1=\naw{-\sqrt{\frac{2}{3}},\frac{1}{\sqrt{3}},0},& v_2=\naw{0,\frac{\sqrt{3}}{2},\frac{1}{2}},\\
v_3=\naw{0,0,1},& v_4=\naw{-\sqrt{\frac{2}{3}},-\frac{1}{2\sqrt{3}},\frac{1}{2}}, \\
v_5=\naw{\sqrt{\frac{2}{3}},-\frac{1}{\sqrt{3}},0},&
v_6=\naw{0,-\frac{\sqrt{3}}{2},-\frac{1}{2}},\\ 
v_7=\naw{0,0,-1}, &v_8=\naw{\sqrt{\frac{2}{3}},\frac{1}{2\sqrt{3}},-\frac{1}{2}},\\
v_9=\naw{0,-\frac{\sqrt{3}}{2},\frac{1}{2}}, &
v_{10}=\naw{0,\frac{\sqrt{3}}{2},-\frac{1}{2}},\\ 
v_{11}=\naw{\sqrt{\frac{2}{3}},\frac{1}{2 \sqrt{3}},\frac{1}{2}}, & 
v_{12}=\naw{-\sqrt{\frac{2}{3}},-\frac{1}{2 \sqrt{3}},-\frac{1}{2}}.\end{array}\end{displaymath}
\end{itemize}
\item The truncated octahedron
\begin{itemize}
\item[] Symmetry group: $O_h$
\item[] Orbit generating group: $S_4$ or $O_h$
\item[] The initial vector: $v_1=\naw{\sqrt{\frac{3}{5}},0,\sqrt{\frac{2}{5}}}$
\item[] The vertices: 
\begin{displaymath}\begin{array}{ll}
v_1=\naw{\sqrt{\frac{3}{5}},0,\sqrt{\frac{2}{5}}}, & v_2=\naw{\sqrt{\frac{3}{5}},\sqrt{\frac{3}{10}},\frac{1}{\sqrt{10}}}, \\ v_3=\naw{\sqrt{\frac{3}{5}},0,-\sqrt{\frac{2}{5}}}, &
v_4=\naw{\sqrt{\frac{3}{5}},\sqrt{\frac{3}{10}},-\frac{1}{\sqrt{10}}},\\
v_5=\naw{\sqrt{\frac{3}{5}},-\sqrt{\frac{3}{10}},\frac{1}{\sqrt{10}}}, & v_6=\naw{\sqrt{\frac{3}{5}},-\sqrt{\frac{3}{10}},-\frac{1}{\sqrt{10}}},\\ 
v_7=\naw{-\sqrt{\frac{3}{5}},0,\sqrt{\frac{2}{5}}}, & v_8=\naw{-\sqrt{\frac{3}{5}},\sqrt{\frac{3}{10}},\frac{1}{\sqrt{10}}}, \\
v_9=\naw{-\sqrt{\frac{3}{5}},0,-\sqrt{\frac{2}{5}}}, &
v_{10}=\naw{-\sqrt{\frac{3}{5}},\sqrt{\frac{3}{10}},-\frac{1}{\sqrt{10}}},\\ v_{11}=\naw{-\sqrt{\frac{3}{5}},-\sqrt{\frac{3}{10}},\frac{1}{\sqrt{10}}}, & v_{12}=\naw{-\sqrt{\frac{3}{5}},-\sqrt{\frac{3}{10}},-\frac{1}{\sqrt{10}}},\\
v_{13}=\naw{-\frac{1}{\sqrt{15}},2 \sqrt{\frac{2}{15}},-\sqrt{\frac{2}{5}}},&
v_{14}=\naw{\frac{1}{\sqrt{15}},\sqrt{\frac{5}{6}},-\frac{1}{\sqrt{10}}},\\
v_{15}=\naw{-\frac{1}{\sqrt{15}},2 \sqrt{\frac{2}{15}},\sqrt{\frac{2}{5}}}, & v_{16}=\naw{\frac{1}{\sqrt{15}},\sqrt{\frac{5}{6}},\frac{1}{\sqrt{10}}},\\
 v_{17}=\naw{\frac{1}{\sqrt{15}},-2 \sqrt{\frac{2}{15}},\sqrt{\frac{2}{5}}}, & v_{18}=\naw{-\frac{1}{\sqrt{15}},-\sqrt{\frac{5}{6}},\frac{1}{\sqrt{10}}},\\
v_{19}=\naw{\frac{1}{\sqrt{15}},-2 \sqrt{\frac{2}{15}},-\sqrt{\frac{2}{5}}}, & v_{20}=\naw{-\frac{1}{\sqrt{15}},-\sqrt{\frac{5}{6}},-\frac{1}{\sqrt{10}}},\\
v_{21}=\naw{-\frac{1}{\sqrt{15}},\frac{1}{\sqrt{30}},\frac{3}{\sqrt{10}}}, & v_{22}=\naw{\frac{1}{\sqrt{15}},-\frac{1}{\sqrt{30}},\frac{3}{\sqrt{10}}},\\ 
v_{23}=\naw{\frac{1}{\sqrt{15}},-\frac{1}{\sqrt{30}},-\frac{3}{\sqrt{10}}}, & v_{24}=\naw{-\frac{1}{\sqrt{15}},\frac{1}{\sqrt{30}},-\frac{3}{\sqrt{10}}}.\end{array}\end{displaymath}
\end{itemize}
\end{enumerate}

\section{Appendix}
In order to describe the threedimensional irreducible representation of $S_4$ group it is sufficient to write out the matrices representing transpositions. They read: 
\begin{equation}
\begin{split}
& D\naw{12}=\left[\begin{array}{ccc}
1 & 0 & 0\\
0 & 1 & 0\\
0 & 0 & -1
\end{array}\right],\qquad
 D\naw{13}=\left[\begin{array}{ccc}
1 & 0 & 0\\
0 & -\frac{1}{2} & -\frac{\sqrt{3}}{2}\\
0 & -\frac{\sqrt{3}}{2} & \frac{1}{2}
\end{array}\right]\\
& D\naw{14}=\left[\begin{array}{ccc}
-\frac{1}{3} & -\frac{\sqrt{2}}{3} & -\frac{\sqrt{6}}{3}\\
-\frac{\sqrt{2}}{3} & \frac{5}{6} & -\frac{\sqrt{3}}{6}\\
-\frac{\sqrt{6}}{3} & -\frac{\sqrt{3}}{6}& \frac{1}{2}
\end{array}\right],\qquad
 D\naw{23}=\left[\begin{array}{ccc}
1 & 0 & 0\\
0 & -\frac{1}{2} & \frac{\sqrt{3}}{2}\\
0 & \frac{\sqrt{3}}{2} & \frac{1}{2}
\end{array}\right]
\\
&D\naw{24}=\left[\begin{array}{ccc}
-\frac{1}{3} & -\frac{\sqrt{2}}{3} & \frac{\sqrt{6}}{3}\\
-\frac{\sqrt{2}}{3} & \frac{5}{6} & \frac{\sqrt{3}}{6}\\
\frac{\sqrt{6}}{3} & \frac{\sqrt{3}}{6}& \frac{1}{2}
\end{array}\right],\qquad
 D\naw{34}=\left[\begin{array}{ccc}
-\frac{1}{3} & \frac{\sqrt{8}}{3} & 0\\
\frac{\sqrt{8}}{3} & \frac{1}{3} & 0\\
0 & 0 & 1
\end{array}\right].\end{split}\label{30}
\end{equation}
The $O_h$ group is the direct product of $S_4\times S_2$; for this reason the threedimensional irreducible representation of this group can be obtained by multiplying all 24 matrices representing $S_4$ by $\pm 1$.

\section{Appendix}
The linear combinations of the coefficients $A_i$ which enter the classical bound according to eq.~\eqref{15h}
\\[0,5cm]
\begin{scriptsize}\begin{tabular}{|c|c|c|c|c|c|c|c|c|c|c|c|c|c|c|c|c|c|c|c|c|c|c|c|c|}
\hline
i & 1&2&3&4&5&6&7&8&9&10&11&12&13&14&15&16&17&18&19&20&21&22&23&24\\
\hline
1&-&0&+&+&0&0&0&0&+&0&0&0&0&+&-&0&-&-&+&0&-&-&0&+\\
\hline
2&-&+&0&0&0&0&+&0&0&+&+&+&0&0&-&+&-&-&0&0&-&-&0&0\\
\hline
3&-&0&0&0&+&+&0&+&0&0&0&0&+&0&-&0&-&-&0&+&-&-&+&0\\\hline
4&0&+&0&-&0&-&-&0&0&0&0&-&+&0&0&+&+&0&-&+&+&0&-&0\\\hline
5&+&0&0&-&+&-&-&+&0&0&0&-&0&+&0&0&0&+&-&0&0&0&-&+\\\hline
6&0&0&+&-&0&-&-&0&+&+&+&-&0&0&+&0&0&0&-&0&0&+&-&0\\\hline
7&0&0&0&0&+&+&0&-&-&+&-&+&-&-&0&-&0&+&0&0&0&+&0&0\\\hline
8&0&0&+&0&0&0&0&-&-&0&-&0&-&-&+&-&0&0&+&+&+&0&+&0\\\hline
9&+&+&0&+&0&0&+&-&-&0&-&0&-&-&0&-&+&0&0&0&0&0&0&+\\\hline
\end{tabular}\end{scriptsize}
\\[0,5cm]
For example, the first combination reads
$$-A_1+A_3+A_4+A_9+A_{14}-A_{15}-A_{17}-A_{18}+A_{19}-A_{21}-A_{22}+A_{24}.$$

\end{document}